\documentclass[sigconf, screen]{acmart}

\copyrightyear{2025}
\acmYear{2025}
\setcopyright{cc}
\setcctype{by}
\acmConference[FSE Companion '25]{33rd ACM International Conference on the
Foundations of Software Engineering}{June 23--28, 2025}{Trondheim, Norway}
\acmBooktitle{33rd ACM International Conference on the Foundations of
Software Engineering (FSE Companion '25), June 23--28, 2025, Trondheim,
Norway}\acmDOI{10.1145/3696630.3728606}
\acmISBN{979-8-4007-1276-0/2025/06}

\usepackage{comment}
\usepackage{microtype}
\usepackage{paralist}
\usepackage{color}
\usepackage[normalem]{ulem}
\usepackage{array}

\newcommand{\Binpool}{\texttt{BinPool}}

\newcommand{\mclearpage}{}

\begin{document}

\title{BinPool: A Dataset of Vulnerabilities for Binary Security Analysis}

\author{Sima Arasteh}
\authornote{Equal contributions.}
\affiliation{
  \institution{University of Southern California}
  \city{Los Angeles}
  \state{CA}
  \country{USA}}
\email{arasteh@usc.edu}

\author{Georgios Nikitopoulos}
\authornotemark[1]
\affiliation{
  \institution{Dartmouth College}
  \city{Hanover}
  \state{NH}
  \country{USA}}
\affiliation{
  \institution{University of Thessaly}
  \city{Volos}
  \state{Thessaly}
  \country{Greece}}
\email{Georgios.Nikitopoulos.GR@dartmouth.edu}

\author{Wei-Cheng Wu}
\affiliation{
  \institution{Dartmouth College}
  \city{Hanover}
  \state{NH}
  \country{USA}}
\email{Wei-Cheng.Wu.GR@dartmouth.edu}

\author{Nicolaas Weideman}
\affiliation{
  \institution{USC Information Sciences Institute}
  \city{Los Angeles}
  \state{CA}
  \country{USA}
}
\email{nhweideman@gmail.com}

\author{Aaron Portnoy}
\affiliation{%
  \institution{Dartmouth College}
  \city{Hanover}
  \state{NH}
  \country{USA}}
\email{aaron.portnoy@dartmouth.edu}

\author{Mukund Raghothaman}
\affiliation{
  \institution{University of Southern California}
  \city{Los Angeles}
  \state{CA}
  \country{USA}}
\email{raghotha@usc.edu}

\author{Christophe Hauser}
\affiliation{
  \institution{Dartmouth College}
  \city{Hanover}
  \state{NH}
  \country{USA}}
\email{christophe.hauser@dartmouth.edu }




\begin{abstract}
The development of machine learning techniques for discovering software vulnerabilities relies
fundamentally on the availability of appropriate datasets. The ideal dataset consists of a large and
diverse collection of real-world vulnerabilities, paired so as to contain both vulnerable and
patched versions of each program. Naturally, collecting such datasets is a laborious and
time-consuming task. Within the specific domain of vulnerability discovery in binary code, previous
datasets are either publicly unavailable, lack semantic diversity, involve artificially introduced
vulnerabilities, or were collected using static analyzers, thereby themselves containing incorrectly
labeled example programs.

In this paper, we describe a new publicly available dataset which we dubbed \Binpool{}, containing numerous samples of
 vulnerable versions of Debian packages across the years. The dataset was automatically curated, and contains both
vulnerable and patched versions of each program, compiled at four different optimization levels.
Overall, the dataset covers 603 distinct CVEs across 89 CWE classes, 162 Debian packages, and
contains 6144 binaries.
We argue that this dataset is suitable for evaluating a range of security analysis
tools, including for vulnerability discovery, binary function similarity, and plagiarism detection.
\end{abstract}

\maketitle
\renewcommand{\shortauthors}{Arasteh et al.}

\mclearpage
\section{Introduction}
\label{sec:intro}

Developing and evaluating software vulnerability detection tools---particularly those that rely on
machine learning~\cite{wang2022jtrans, ding2019asm2vec, grieco2016toward, pewny2015cross,arakelyan2021bin2vec, eschweiler2016discovre, ji2021buggraph, xu2017neural, arasteh2024:acsac}---%
fundamentally depends on the existence of robust, labeled datasets of known vulnerabilities.
Assembling these datasets is often among the hardest parts of building these 
systems.

We argue that there are severe limitations with existing datasets:
\emph{First}, well-annotated and easy-to-compile datasets such as Juliet~\cite{meade2012juliet} are
limited to small programs and artificially introduced bugs. The lack of diverse, real-world code
examples leads to fears of overfitting and limited generalization in trained models.
\emph{Next}, many previous papers do not make their data publicly available~%
\cite{yunlong2019genius, eschweiler2016discovre, gao2018vulseeker, gao2018vulseekerpro}.
\emph{In addition}, many publicly available datasets focus on binary similarity detection,
rather than on vulnerability finding~\cite{wang2022jtrans}.
\emph{Finally}, some datasets such as those assembled by Pereira et al.~\cite{Pereira:2022}
rely on warnings reported by program analysis tools such as Flawfinder~%
\cite{wheelerflawfinder} and Cppcheck~\cite{marjamaki2007cppcheck}. Naturally, this results in
possibly tainted ground truth, leading to fears of incorrectly trained classifiers.

In contrast, the ideal dataset contains a large, diverse collection of programs with different and
well-annotated vulnerability types. These programs should represent examples of real-world
vulnerabilities, and have a matched instances of vulnerable and patched code. Furthermore,
especially for vulnerability discovery at binary level, it is important to be able to compile code into
executable files.

In this paper, we report on a new dataset of binaries with historical real-world vulnerabilities
named \Binpool. This is a collection of high-impact vulnerabilities obtained by crawling and
collating the National Vulnerability Database (NVD)~\cite{nvdNist2023} with data from the Debian
security tracker~\cite{debian_security_tracker} and the archive of Debian snapshots~%
\cite{debianSnapshot2023}. We have already used an early version of this dataset to evaluate BinHunter~\cite{arasteh2024:acsac}, a graph neural network-based binary vulnerability detection system [2]. Going forward, we believe that this dataset will be very useful for practical security research.

At a high level, we assembled the dataset as follows: We mapped each CVE in the NVD database to its
associated CWE, Debian packages that fix the vulnerability, and the specific functions and lines
that were edited as part of the patch. We then wrote scripts to automatically build these Debian
packages. By using the Debian package maintenance tool \texttt{quilt} to selectively apply and
withhold the patch from the source code, we obtained vulnerable and patched versions of each binary.
We then repeated this process at four different optimization levels. The entire process is highly
automated, and currently contains 603 distinct CVEs spanning 89 different CWE categories, across 162
Debian packages, and with 6144 binaries in total. Overall, the dataset spans 910 source functions
and 7280 binary functions respectively.

We note that \Binpool{} can be used for both vulnerability discovery and binary function similarity
detection. In the case of vulnerability discovery, it is applicable to both source and binary code.
It includes a range of vulnerabilities, from highly specific categories such as CWE-122 (Stack-based
buffer overflow) to wider categories such as CWE-119 (Improper restriction of operations with the
bounds of a memory buffer). Notably, the dataset contains detailed information about the precise
location of each vulnerability, including files, functions, and lines affected by the patch, both at
the source and binary levels. The automation scripts and links to the dataset may found at~%
\url{https://github.com/SimaArasteh/binpool}.

The rest of this paper is organized as follows:
We describe the \Binpool{} collection process in
Section~\ref{sec:Dataset-Construction},
the overall structure of the dataset in Section~%
\ref{sec:Dataset-structure}, and
its potential applications in Section~\ref{sec:application}
respectively. In Sections~\ref{related-work} and~\ref{limitations} we describe the related work and
discuss the limitations of our approach.
        \mclearpage
\section{Dataset Construction}
\label{sec:Dataset-Construction}

\begin{figure*}
\centering
\includegraphics[width=.95\textwidth]{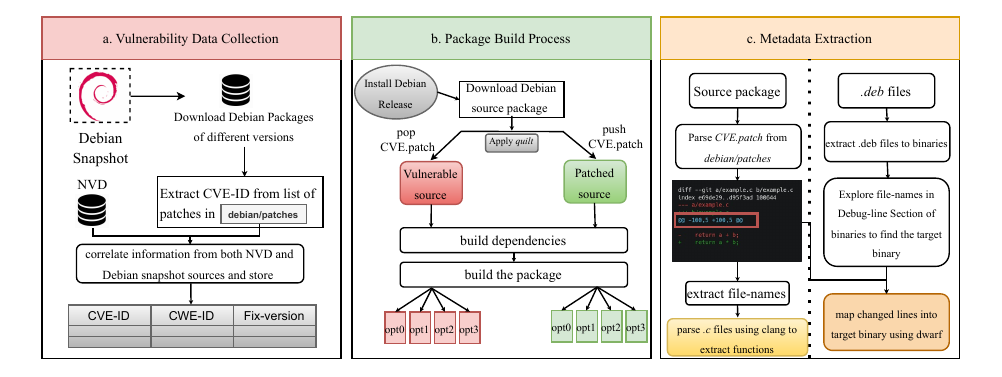}
\caption{Construction process of the \Binpool{} dataset.
  (\textit{a}) In the first phase, we collect data about the vulnerabilities by gathering CVE-IDs,
  CWEs, and the affected and fixed package versions from the Debian snapshots and NVD databases.
  (\textit{b}) In the second phase, we build packages for both vulnerable and patched versions.
  (\textit{c}) In the last phase, we extract detailed metadata, including function names and
  vulnerability locations (both at source and binary levels).}
\label{fig:workflow}
\end{figure*}

We show the workflow of the \Binpool{} curation process in Figure~\ref{fig:workflow}. Its
construction is enabled by three resources provided by the Debian project:
First, Debian packages are structured with control files, metadata and content, and are maintained
by the community through updates and security patches.
Second, Debian Snapshots~\cite{debianSnapshot2023} contains an archive of historical package
versions, allowing access to specific releases over time.
Finally, the Debian Security Tracker~\cite{debian_security_tracker} monitors vulnerabilities such as
CVEs, enabling users and maintainers to stay informed about security issues. We use the
beautifulsoup library to crawl data from these resources.

\subsection{Vulnerability Data Collection}
\label{subsec:vulcollect}

The Debian Security Tracker~\cite{debian_security_tracker} provides a frequently updated JSON file
with CVE-IDs and version information about the fixed packages.%
\footnote{\url{https://security-tracker.debian.org/tracker/data/json}} We gather the relevant
version numbers and recover the corresponding CWE categories by consulting the NVD database~%
\cite{nvdNist2023}. We then use the Debian Snapshots archive to gather a link to the package source
code. We record this information in a publicly available \href{https://docs.google.com/spreadsheets/d/1qztIwB8xJ10H-2HLX15vI29Ze7yFDOrv7kDQ4JUi1g8/edit?usp=sharing}{Google Spreadsheet}.


\subsection{Package Build Process}
\label{subsec:built}

Next, Debian offers an automatic system for building packages using the \texttt{build-dep} and
\texttt{dpkg-buildpackage} tools. The \texttt{build-dep} tool installs the necessary dependencies
and the \texttt{dpkg-buildpackage} tool automatically compiles Debian packages from their source
code. Our automation system leverages these Debian tools to fully streamline the package building
process.

We retrieve source code of the fixed versions of each package from Debian Snapshots. Each package
includes a directory, \texttt{debian/patches}, which contains a sequential list of patches to be
applied to the source code. We use the \texttt{quilt} tool to selectively apply or remove the
specific patches that fixed the vulnerability. Finally, we build each variant (buggy / patched) of
the package at four different optimization levels, including with debug symbols.


\subsection{Metadata Extraction}
\label{subsec:metadata}

The last conceptual step is to extract metadata information from these packages. This information
includes files, functions, and source and binary locations that were modified by the patch.

The package build process results in a number of \texttt{deb} files being produced as output.
Informally, \texttt{deb} files are used to distribute and install Debian packages. We extract and
search through these packages to locate the individual binaries (ELF files) that were affected by
the patch.
Now: to extract the metadata, we parse the patch file and recover the files and modules that were
edited as part of fixing the vulnerability. We then parse the modules in the Debian source using
the \texttt{clang} compiler front-end to extract the relevant function names. Finally, we obtain
file names, function names and lines modified as part of the patch.

In order to find the target binary among extracted \texttt{deb} files, we map file names in the
patch into the corresponding compilation units in the binary using the debug line section in DWARF.
Recall that each compilation unit includes a source file and associated headers that are compiled
together to produce a single object file. This mapping process allows us to determine which binary
contains the patched file or compilation unit. After finding the target binary, we once again use
debug information embedded in the binary to extract the precise memory offsets corresponding to the
lines of code modified in the patch. All steps in this dataset construction pipeline are fully
automated and written in Python.
 \mclearpage
\section{Structure of the BinPool Dataset}
\label{sec:Dataset-structure}
\begin{figure}
\centering
\includegraphics[width=0.35\textwidth]{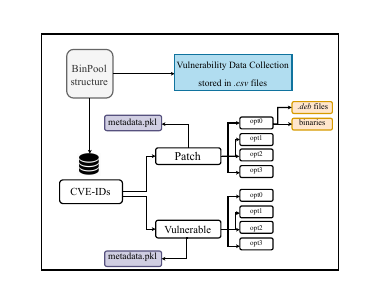}
\caption{Structure of the \Binpool{} dataset. The central CSV file contains information about CVE
  and CWE-IDs, version numbers, and links to source code. In parallel, the dataset is organized
  according to the vulnerability IDs. Each vulnerability includes metadata about the function names,
  module names, and affected code locations in \texttt{pkl} files, and versions of the vulnerable
  and patched \texttt{deb} files and binaries obtained from different optimization levels.}
\label{fig:structure}
\end{figure}

There are three principal components in \Binpool{} dataset: The metadata (stored in \texttt{pkl}
files), the binaries in question, and a central CSV file containing information about CVEs, CWEs,
version numbers, and links to source code. We also gather all the metadata in a JSON file called \texttt{binpool\_info.json}. The links to these resources may be found in the main
artifact repository: \url{https://github.com/SimaArasteh/binpool}. We show the layout of these
directories in Figure~\ref{fig:structure}. To the best of our knowledge, \Binpool{} is the first such
dataset to provide this level of detailed information. We also present some aggregate statistics
about the dataset in Table~\ref{tab:measurements}.

\begin{table}
\caption{Aggregate statistics about \Binpool.}
\label{tab:measurements}
\centering
\begin{tabular}{|c|c|}
\hline Measurement & Value \\
\hline Number of unique CVEs & 603 \\
\hline Number of CWEs & 89 \\
\hline Number of Debian packages & 162 \\
\hline Total number of binaries & 6144 \\
\hline Total number of source modules & 768 \\
\hline Total number of source functions & 910 \\
\hline Total number of binary functions & 7280 \\
\hline
\end{tabular}
\end{table}
    \mclearpage
\section{Possible Applications}
\label{sec:application}

The immediate intended application of \Binpool{} is in the development and evaluation of
vulnerability detection techniques. We expect the collection of functions with diverse real-world
semantics to form a challenging dataset and for the framework to be an ongoing benchmark collection
technique for bug-finding tools. An early smaller-scale version of the dataset already formed a
signficant part of our evaluation methodology for BinHunter~\cite{arasteh2024:acsac}.

Beyond just machine learning techniques, we expect the dataset to be useful for benchmarking other
program analysis systems such as angr~\cite{shoshitaishvili2016sok}. We also note that many vulnerability detection
tools are based on sophisticated reasoning pipelines involving data flow and control flow analysis
and information about types of vulnerabilities. As such, many of these intermediate analyses---such
as precise inter-procedural data flow analyses in binaries---are of independent research interest,
and we expect \Binpool{} to also form a possible benchmarks for these applications.

Finally, beyond just vulnerability discovery, we expect \Binpool{} to be useful in other binary
analysis problems. As one example, recall that we provide multiple versions of each binary package,
compiled using different optimization levels. These matched binaries might form a good benchmark for
code search and function similarity detection algorithms.

\mclearpage
\section{Related Work}
\label{related-work}

In this section, we review the history of vulnerability datasets at both the source code and binary
levels. We compare BinPool features with state-of-the-art datasets for both source code and binaries with different applications in Table~\ref{tab:comparison}.

\begin{table}
\centering
\caption{A comparison between existing
datasets and BinPool. VD, CP and BSD stand for vulnerability discovery, compiler provenance and binary similarity detection respectively.}
\label{tab:comparison}
\scalebox{0.6}{
\begin{tabular}{|>{\centering\hspace{0pt}}m{0.2\linewidth}|>{\centering\hspace{0pt}}m{0.2\linewidth}|>{\centering\hspace{0pt}}m{0.2\linewidth}|>{\centering\hspace{0pt}}m{0.304\linewidth}|>{\centering\hspace{0pt}}m{0.17\linewidth}|>{\centering\arraybackslash\hspace{0pt}}m{0.17\linewidth}|} 
\hline
Dataset    & source-level & binary-level & Application                                               & \# CVEs & \# projects                               \\ 
\hline
CrossVul   & Yes          & -            & VD                                   & 5131   & 1675                                     \\[1ex]
ReposVul   & Yes          & -            & VD                                   & 6,134  & 1,491                                    \\[1ex]
BigVul     & Yes          & -            & VD                                  & 3754   & 348                                      \\[1ex]
MegaVul    & Yes          & -            & VD                                   & 8,254  & 992                                      \\[1ex]
BinBench   & -            & Yes          & BSD, CP      & -      & 131                                      \\[1ex]
BinaryCorp & -            & Yes          & BSD                              & -      & 9819                                     \\[1ex]
BinKit     & -            & Yes          & BSD                               & -      & 51                                       \\[1ex]
\hline
\textbf{BinPool}    & -            & Yes          & BSD,VD & 603    & \textcolor[rgb]{0.122,0.137,0.157}{162}  \\
\hline
\end{tabular}
}
\end{table}


\subsection{Source-Level Datasets}

In part because of their relative ease of collection, most existing datasets of vulnerable code are
at the source level. One particularly famous example is the Juliet dataset~\cite{meade2012juliet},
which (among other languages) provides a collection of C and C++ programs with artificially injected
vulnerabilities. The dataset provides macros to switch between vulnerable and non-vulnerable
versions of code. As such, because of its scale and ease of use, it has been used to train and
evaluate numerous program analysis tools.

Of course, one important complaint against the Juliet dataset is that it includes artificial, as
opposed to real-world examples of bugs. One convenient approach to do this (and similar to our
approach in \Binpool) is to use the NVD dataset to identify examples of actual vulnerabilities in
production open-source code~\cite{nikitopoulos2021crossvul}. As such, this provides a reliable
indicator of ground truth. Another alternative to create these datasets~\cite{lipp2022empirical} is
to identify vulnerabilities using static analysis tools such as Cppcheck~%
\cite{marjamaki2007cppcheck} and Flawfinder~\cite{wheelerflawfinder}.

Another aspect of datasets is the granularity of information provided about the location of the
vulnerability. For example, datasets such as Bigvul~\cite{fan2020ac}, Megavul~\cite{ni2024megavul},
and Crossvul~\cite{nikitopoulos2021crossvul} provide large collections of vulnerable source code
with precise information about the software fault. These are typically extracted by using commit
information. Crossvul covers over 40 programming languages, whereas Bigvul and Megavul focus
specifically on vulnerabilities in C/C++ programs. Compared to Bigvul, Megavul encompasses more
vulnerabilities and open-source projects, and it includes information on vulnerable functions using
a Tree-sitter parser. While Crossvul is more diverse in covering multiple programming languages, it
lacks detailed function information.

Among these datasets, Reposvul stands out by offering a repository-level dataset. This dataset aims
to address three issues in existing collections. First, many patches are not strictly
security-related and are mixed with non-security changes. To address this, Reposvul uses large
language models (LLMs) and static analysis tools to identify security-specific patches. Second, most
datasets focus only on function-level vulnerabilities, overlooking the importance of
inter-procedural vulnerability analysis. Reposvul addresses this by capturing relationships between
functions involved in a patch. Lastly, it identifies outdated patches by tracking commit histories.


\subsection{Binary-Level Datasets}

While there are many diverse source-level vulnerability datasets, binary-level datasets are limited.
The main challenge is that compiling programs from source code frequently requires considerable
manual effort. Our solution to this challenge in \Binpool{} is to use the existing build system
supplied as part of Debian to obtain a scalable and well-tested build system for a large repository
of packages.

Existing binary-level vulnerability datasets face several challenges. Firstly, many of these datasets are not publicly accessible. Secondly, most are derived from a small set of open-source projects, compiled across various optimization levels and architectures. These datasets are primarily designed for detecting similarities between different architectures and optimization levels, rather than focusing specifically on vulnerability discovery. Additionally, the limited range of projects they include can lead to overfitting in machine learning models.

For example, the Genius dataset~\cite{feng2016scalable} includes only 154 vulnerable functions sourced from BusyBox, OpenSSL, and Coreutils, compiled across three architectures, four optimization levels, and two compilers. The Vulseekerpro dataset~\cite{gao2018vulseeker} features only 15 unique CVEs, compiled with different optimization settings. Among the various datasets of binary programs, Jtrans~\cite{wang2022jtrans} is the most diverse, featuring programs compiled from multiple projects. However, it is specifically developed for binary similarity detection and is limited to identifying particular CVEs across different optimization levels.
      \mclearpage
\section{Limitations}
\label{limitations}


Although our dataset includes a diverse and comprehensive collection of real-world
vulnerabilities with detailed metadata, it still has only a number of CVEs per CWE category. As a
result, although the dataset is ideal for \emph{evaluating} vulnerability detection tools, it might
be insufficient for \emph{training} classifiers. As a consequence, the training process might need
to be supplemented with other sources of information, including datasets such as Juliet, in a manner
similar to our work on BinHunter~\cite{arasteh2024:acsac}.

A second limitation of \Binpool{} is that it was collected purely using information about
modifications to source code, and therefore does not include some important sources of information
about the errors in question. For example, one might be able to augment the data with examples of
failing test cases, error traces resulting from symbolic execution engines such as angr~\cite{shoshitaishvili2016sok},
or include more detailed information about inter-procedural data flows. Of course, collecting such
information would require significant extensions to our data collection pipeline, but might
conceivably lead to more sophisticated vulnerability detection tools.

  \mclearpage
\section{Conclusion}
\label{concolusion}

In this paper, we have introduced a public dataset of historical vulnerabilities in Debian packages.
The dataset provides detailed information, including CVE and CWE identifiers, version numbers, and
lists of the functions, files, and lines modified as part of the fix, at both source and binary
levels. The dataset is suitable for both developing and evaluating both vulnerability discovery and
binary function similarity tools. In particular, we envision \Binpool{} to be appropriate as a test
set for evaluating machine learning-based techniques, and can potentially also be utilized with
program analysis tools such as angr.

We intend to continuously run the data collection pipeline and grow the dataset over time. We also
plan to include more detailed information, such as the results of inter-procedural data flow
analyses, thereby capturing how vulnerabilities arise from the interaction of multiple functions
within each program. We hope that this information leads to more effective and better evaluated
techniques for vulnerability discovery.

        \mclearpage

\bibliographystyle{plain}
\bibliography{references}

\end{document}